\documentclass[11pt,a4paper]{article}
\usepackage[utf8]{inputenc} 

\usepackage{wrapfig,float,floatflt}
\usepackage{amsmath,amssymb,amscd,multicol,amsfonts,mathabx, textcomp}
\usepackage{amsthm}
\usepackage{graphicx,stmaryrd}
\usepackage[matrix,arrow,curve]{xy}
\usepackage{colortbl}
\usepackage{floatflt}
\usepackage{appendix}
\begin{document}
\title{\bf Unusual square roots\\ in the ghost-free theory of massive gravity}

\author{Alexey Golovnev${}^{a,b}$,\quad  Fedor Smirnov${}^{a}$\\
{${}^{a}$\it Faculty of Physics, St. Petersburg State University;}\\ 
{\it Ulyanovskaya ul., d. 1, Saint Petersburg 198504, Russia,}\\
{${}^{b}$\it St Petersburg National Research University of Information Technologies, Mechanics and Optics;}\\ 
{\it Kronverkskiy pr. 49, Saint Petersburg, 197101, Russia}\\
{\small agolovnev@yandex.ru} \qquad {\small  fsmirnov94@gmail.com}}
\date{}

\maketitle

\begin{abstract}

A crucial building block of the ghost free massive gravity is the square root function of a matrix. This is a problematic entity from the viewpoint of existence and uniqueness properties. We accurately describe the freedom of choosing a square root of a (non-degenerate) matrix. It has discrete and (in special cases) continuous parts. When continuous freedom is present, the usual perturbation theory in terms of matrices can be critically ill defined for some choices of the square root. We consider the new formulation of massive and bimetric gravity which deals directly with eigenvalues (in disguise of elementary symmetric polynomials) instead of matrices. It allows for a meaningful discussion of perturbation theory in such cases, even though certain non-analytic features arise.

\end{abstract}
\section{Introduction}

Over the last century the theory of general relativity has been confronted with experiment at ever increasing precision in a big variety of scales. It works very well in many different circumstances, from Solar System tests and even GPS technology to cosmological perturbation theory. However, in the latter case there is a number of issues that might hint towards modifications of GR. Those include the missing mass and the accelerated expansion problems. It perfectly motivates a whole variety of approaches to modify or extend the theory of gravitational interactions. Massive (and bimetric) gravity represent a rather popular option for that; one which is theoretically challenging but promising and relatively well studied.

The idea of a massive graviton dates back to classical work of Fierz and Pauli in 1939 $\cite{FP}$ which at the time was more of a theoretical excercise. At later times, the naive expectation was that the Yukawa exponent should give rise to an effective cut-off for the gravitational force on large scales leading to possible explanation of the Dark Energy phenomenon. In reality it turned out not that simple, of course. And, even at the most formal level, one of the earliest issues was the infamous vDVZ discontinuity (van Dam-Veltman-Zakharov \cite{MMYM,LGT}) discovered in 1970: when the graviton mass tends to zero, the (by definition, linearised) Fierz-Pauli theory does not restore massless GR since the scalar mode does not completely decouple. This problem can have a resolution via non-linear effects known under the name of Vainshtein mechanism \cite{VM} which has been proposed in 1972, see also Ref. \cite{IVM}.

However, the second significant problem was precisely about building a non-linear realisation of the Fierz-Pauli model. In GR ten independent variables are reduced to two degrees of freedom because the lapse and shifts play the role of four Lagrange multipliers enforcing four first class constraints in the spatial sector of four otherwise independent variables. If one modifies the theory, the non-dynamical variables generically would enter the action non-linearly leading to disappearance of constraints in the spatial sector and to six degrees of freedom being left. Five of them are the usual polarisations of massive spin two, while the remaining one is a ghost, the Boulware-Deser ghost \cite{BD}. Therefore, a modified theory needs to have a constraint for elimination of the ghost. A beautiful solution was suggested in 2010 by de Rham, Gabadadze and Tolley \cite{GFPA, RMG}. It has a pair of second class constraints for the spatial sector \cite{NLAMG,RGPMG}. 

Obviously, such models require a second metric on top of the physical one (in order to have an invariant non-derivative interaction potential, in the first place). Initially, the model made use of Minkowski metric $\eta_{\mu\nu}$ which serves as a proper background for perturbation theory in the weak gravity limit. However, it was quickly generalised to an arbitrary non-dynamical symmetric tensor $f_{\mu \nu}$ \cite{GFMGGRM, CSCAG}, and even to the full-fledged bimetric gravity \cite{CSCAG, BGGMG} with two dynamical metrics $g_{\mu \nu}$ and $f_{\mu \nu}$. Massive gravity and its various extensions have been widely applied in cosmological settings. Having a viable cosmology with massive gravity appears non-trivial \cite{MasCo} but not impossible \cite{CS, Fawad}. Actually, it might even be good enough for reproducing effects of Dark Matter \cite{HSDM}. Nice reviews of massive and bimetric gravity can be found in Refs. \cite{MG, RDBT}.

As we have already stated above, massive gravity has two metrics, one of which ($g_{\mu\nu}$) is the physical one which couples to matter, while the second one is an auxiliary (fiducial) metric which can either be fixed of dynamical. They go with an interaction potential made of $\sqrt{g^{-1}f}$, the square root of the matrix ${\mathbb X}^{\mu}_{\nu}\equiv g^{\mu\alpha}f_{\alpha\nu}$. Among very many papers devoted to different aspects of bimetric gravity, only very few \cite{Cedric, GHW1, GHW2, NBMG, ADMAMG} deal with the formal aspects of solving the matrix equation $\mathbb A^2 = \mathbb X$ that arises in theory and defines the interaction between metrics. This paper aims at filling this yawning gap based on our new approach \cite{Dealing17} which deals with eigenvalues instead of matrix components.

The paper is organised as follows. In Section 2 we review the bimetric gravity. In Section 3 (and Appendix A) we describe square roots of matrices and related problems of massive gravity. In Section 4 we describe our new formulation of the model. In Section 5 we apply it to perturbation theory around unusual square roots with continuous ambiguity. Finally, in Section 6 we discuss implications and conclude.

\section{Bimetric theory}
\label{Bimetric theory}

Let us first review the basic foundations of bimetric theory. The action can be written (in four dimensions) as

\begin{multline}
\label{TA}
S=-\frac{M_g^2}{2} \int d^4 x \sqrt{-g} R_g-\frac{M_f^2}{2} \int d^4 x \sqrt{-f} R_f \\
+m^2 M_g^2 \int d^4 x \sqrt{-g} \sum_{n=0}^4 \beta_n e_n\left(\sqrt{g^{-1}f}\right)+\int d^4 x \sqrt{-g} \mathcal L_m(g, \Phi)
\end{multline}
with the elementary symmetric polynomials $e_n\equiv\sum\limits_{i_1<i_2<\ldots<i_n}\lambda_{i_1}\lambda_{i_2}\cdots\lambda_{i_n}$ of the eigenvalues $\lambda_i$ of the matrix $\sqrt{g^{-1}f}\equiv\sqrt{\mathbb X}$. By definition we take $e_0(\mathcal M) =  1$, and then the following recurrent relation can be proven:
\begin{equation}
\label{recurrent}
e_n(\mathcal M)=\frac{1}{n}\sum\limits_{i=1}^n (-1)^{i-1}[\mathcal M^i]\cdot e_{n-i}(\mathcal M)
\end{equation}
for any matrix $\mathcal M$ where $\left[\mathcal M\right]$ denotes the trace of  $\mathcal M$. In particular,
\begin{eqnarray*}
e_1(\sqrt{\mathbb X}) & = & \left[\sqrt{\mathbb X}\right],\\ 
e_2(\sqrt{\mathbb X}) & = & \frac{1}{2}\left(\left[\sqrt{\mathbb X}\right]^2-\left[\sqrt{\mathbb X}^2\right]\right),\\
e_3(\sqrt{\mathbb X}) & = & \frac{1}{6}\left(\left[\sqrt{\mathbb X}\right]^3-3\left[\sqrt{\mathbb X}\right]\cdot\left[\sqrt{\mathbb X}^2\right]+2\left[\sqrt{\mathbb X}^3\right]\right),\\ 
e_4(\sqrt{\mathbb X}) & = &  \frac{1}{24}\left(\left[\sqrt{\mathbb X}\right]^4-6\left[\sqrt{\mathbb X}\right]^2\left[\sqrt{\mathbb X}^2\right]+3\left[\sqrt{\mathbb X}^2\right]^2+8\left[\sqrt{\mathbb X}\right]\left[\sqrt{\mathbb X}^3\right]-6\left[\sqrt{\mathbb X}^4\right]\right),
\end{eqnarray*}
and in four dimensions we automatically have 
$$e_4(\mathcal M)=\mathrm{det}\left(\mathcal M\right)$$
and $e_n=0$ for $n>4$.
The field $\Phi$ stands in the action (\ref{TA}) for unspecified matter fields which couple to the physical metric $g_{\mu\nu}$. If needed, the number of spacetime dimensions can be changed straightforwardly.

The mass parameter $m$ is redundant with parameters $\beta_i$, where $\beta_0$ and $\beta_4$ (or $\beta_D$ in $D$ dimensions) give rise only to cosmological constants for metrics $g_{\mu \nu}$ and $f_{\mu \nu}$ respectively while three ($D-1$) remaining parameters correspond to three ($D-1$) different interaction terms. The action is invariant under the following simultaneous transformations:
\begin{equation}
\label{gfsymmetry}
g \leftrightarrow f; \phantom{asd} \beta_n \leftrightarrow \beta_{4-n}; \phantom{asd} M_g \leftrightarrow M_f; \phantom{asd} m^2 \leftrightarrow m^2\dfrac{M_g^2}{M_f^2}.
\end{equation}

In order to derive the equations of motion, let us notice that from $\delta \mathbb X = \delta \left(\sqrt{\mathbb X}\cdot \sqrt{\mathbb X}\right)$ follows $\delta \mathbb X = \left(\delta \sqrt{\mathbb X}\right) \cdot\sqrt{\mathbb X} + \sqrt{\mathbb X} \cdot\left(\delta \sqrt{\mathbb X}\right)$, or equivalently $\sqrt{\mathbb X}^{-1}\cdot \delta \mathbb X = \sqrt{\mathbb X}^{-1}\cdot \left(\delta \sqrt{\mathbb X}\right)\cdot \sqrt{\mathbb X} + \delta \sqrt{\mathbb X}$. Taking the trace of this equation we get $\left[\delta \sqrt{\mathbb X}\right] = \dfrac12 \left[\sqrt{\mathbb X}^{-1} \cdot\delta \mathbb X\right]$. More generally, we have the following identity \cite{NLAMG}:
\begin{equation*}
\delta\left[\Big(\sqrt{g^{-1}f}\Big)^{n}\right]=\frac{n}{2} \cdot\left[g \Big(\sqrt{g^{-1}f}\Big)^n \delta g^{-1}\right]
\end{equation*}
which gives
\begin{equation}
\label{deltae}
\frac{2}{\sqrt{-g}} \delta \left(\sqrt{-g} e_n\left(\sqrt{g^{-1}f}\right)\right)=\sum^{n}_{m=0} (-1)^{m+1} \left[g \Big(\sqrt{g^{-1}f} \Big)^m \delta g^{-1}\right]\cdot e_{n-m}\left(\sqrt{g^{-1}f}\right).
\end{equation}

Varying the action (\ref{TA}) with respect to $g_{\mu \nu}$ we obtain by virtue of (\ref{deltae}):
\begin{equation}
\label{eqg}
\stackrel{g\phantom{a.}}{R_{\mu \nu}}-\frac{1}{2}\stackrel{\phantom{.}g}{R}g_{\mu \nu}+\frac{m^2}{2}\sum^3_{n=0}(-1)^n \beta_n \left(\vphantom{\int}g_{\mu \lambda}{\mathbb Y_{(n)}}^{\lambda}_{\nu}\left(\sqrt{g^{-1}f}\right)+g_{\nu \lambda}{\mathbb Y_{(n)}}^{\lambda}_{\mu}\left(\sqrt{g^{-1}f}\right)\right)=\frac{1}{M^2_{g}}T^{\mu}_{\phantom{a} \nu}
\end{equation}
where $\mathbb Y_{(n)}$ are defined as:
$$\mathbb Y_{(0)}\left(\sqrt{\mathbb X}\right)  =  \mathbb I,$$ 
$$\mathbb Y_{(1)}\left(\sqrt{\mathbb X}\right)  =  \sqrt{\mathbb X} - \mathbb{I}\cdot\left[\sqrt{\mathbb X}\right],$$
$$\mathbb Y_{(2)}\left(\sqrt{\mathbb X}\right)  =  {\sqrt{\mathbb X}}^2-\sqrt{\mathbb X}\cdot \left[\sqrt{\mathbb X}\right]+\frac{1}{2}\mathbb{I}\cdot\left(\left[\sqrt{\mathbb X}\right]^2-\left[{\sqrt{\mathbb X}}^2\right]\right),$$
\begin{multline*}
\mathbb Y_{(3)}\left(\sqrt{\mathbb X}\right)  =  {\sqrt{\mathbb X}}^3 - {\sqrt{\mathbb X}}^2\cdot\left[\sqrt{\mathbb X}\right]+\frac{1}{2}\sqrt{\mathbb X}\cdot\left(\left[\sqrt{\mathbb X}\right]^2-\left[\sqrt{\mathbb X}^2\right]\right)\\
-\frac{1}{6}\mathbb{I}\cdot\left(\left[\sqrt{\mathbb X}\right]^3-3\left[\sqrt{\mathbb X}\right]\cdot\left[{\sqrt{\mathbb X}}^2\right]+2\left[{\sqrt{\mathbb X}}^3\right]\right),
\end{multline*}
or in general
\begin{equation}
\label{Y-s}
\mathbb Y_{(n)}\left(\sqrt{\mathbb X}\right)  = \sum_{i=0}^n (-1)^i e_i\left(\sqrt{\mathbb X}\right)\cdot \left(\sqrt{\mathbb X}\right)^{n-i}.
\end{equation}

Analogously, varying the action (\ref{TA}) with respect to $f_{\mu \nu}$, one can get
\begin{equation}
\label{eqf}
\stackrel{f\phantom{a.}}{R_{\mu \nu}}-\frac{1}{2}\stackrel{\phantom{.}f}{R} f_{\mu \nu}+\frac{m^2}{2M^2_{*}}\sum^3_{n=0}(-1)^n \beta_{4-n}\left(\vphantom{\int}f_{\mu \lambda}{\mathbb Y_{(n)}}^{\lambda}_{\nu}\left(\sqrt{f^{-1}g}\right)+f_{\mu \lambda}{\mathbb Y_{(n)}}^{\lambda}_{\mu}\left(\sqrt{f^{-1}g}\right)\right)=0
\end{equation}
where the dimensionless ratio of two Planck masses has been introduced: $M_*=\frac{M_f}{M_g}$. The quantities with letter $g$ (respectively $f$) on top are defined with respect to the metric $g_{\mu\nu}$ (respectively $f_{\mu\nu}$).

For completeness, let us mention that, as a consequence of the Bianchi identity and the covariant conservation of $T_{\mu \nu}$, the $g_{\mu \nu}$ equation of motion (\ref{eqg}) leads to the Bianchi constraint
\[
\stackrel{g}{\nabla}_{\mu} \sum^3_{n=0}(-1)^n \beta_n\left(\vphantom{\int}{\mathbb Y_{(n)}}^{\mu}_{\nu}\left(\sqrt{g^{-1}f}\right)+g^{\mu \alpha}g_{\nu \lambda}{\mathbb Y_{(n)}}^{\lambda}_{\alpha}\left(\sqrt{g^{-1}f}\right)\right)=0.
\]
Similarly for $f_{\mu \nu}$ we find from eq. (\ref{eqf}) that
\[
\stackrel{f}{\nabla}_{\mu} \sum^3_{n=0}(-1)^n \beta_{4-n}\left(\vphantom{\int}{\mathbb Y_{(n)}}^{\mu}_{\nu}\left(\sqrt{f^{-1}g}\right)+f^{\mu \alpha}f_{\nu \lambda}{\mathbb Y_{(n)}}^{\lambda}_{\alpha}\left(\sqrt{f^{-1}g}\right)\right)=0.
\]

In this paper we consider the massive gravity regime with the auxiliary metric $f_{\mu\nu}= \eta_{\mu\nu}$, and therefore the $R_f$ term in the action should be ignored. However, the issues we discuss are generic for any situations in massive and bimetric gravity in metric formulation.

\subsection{The constraint analysis}

We are interested in how far the good properties of the model, including the absence of Boulware-Deser ghost, can go with unusual square roots. Initially, the degrees of freedom count has been done in perturbation theory around the trivial square root of unit matrix. However, it left open the question of self-consistency beyond perturbative level which required methods of non-perturbative Hamiltonian analysis.

The latter was first done by Hassan and Rosen in a series of papers starting from \cite{RGPMG} with the simplest case  of $\beta_1$ model and Minkowski fiducial metric which is enough for the purposes of our discussion. They used the ADM decomposition for $g_{\mu\nu}$:
\begin{equation*}
\mathbb X\equiv g^{\mu \alpha} \eta_{\alpha \nu} = \dfrac{1}{N^2}\begin{pmatrix}
	1 & N^l \delta_{l j} \\
    -N^{j} & (N^2 \gamma^{i l} - N^i N^l)\delta_{l j}
\end{pmatrix}
\end{equation*}
where the lapse $N$, the shift $N_i$, and the spatial metric $\gamma_{ij}$ are the usual ADM variables for $g_{\mu\nu}$, and $\delta_{ij}$ is the Euclidean spatial metric (as a part of fiducial Minkowski $\eta_{\mu\nu}$).

The first order Lagrangian of the $\beta_1$ model can be written in these variables as:
\[
\mathcal{L}^{\text{dRGT}}_{\beta_1} = \pi^{ij} \partial_{t} \gamma_{ij} + N R^0 + N^i R_i + 2 m^2 \sqrt{\gamma} N \left(3 - \left[\sqrt{\mathbb X}\right]\right)
\]
where the number $3$ in the potential stands solely for compensating the effective cosmological constant from the $\beta_1$ term. $R^0$, $R_i$ are the usual constraints of GR, and $\pi^{i j}$ is the canonical conjugate momentum of $\gamma_{i j}$. The crucial feature of generic massive gravity is that the lapse and shift enter the potential term non-linearly, and therefore they fail to serve as Lagrange multipliers. Their variations give algebraic equations for themselves instead of constraining the spatial sector. That's how we get $6$ degrees of freedom: $5$ of massive spin two, and $1$ of the scalar ghost.

A way out would be to have a very special potential which allows one (primary) constraint to survive in the spatial sector. Hassan and Rosen have shown that one can make a (linear in lapse) redefinition of the shift vector
\begin{equation}
\label{Nn}
N^i=(\delta^i_j + ND^i_{\phantom{.}j}(n,\gamma))\cdot n^j
\end{equation}
such that the Lagrangian density of the $\beta_1$ model is rendered linear in the lapse,
\begin{equation*}
N \sqrt{\mathbb X} = \mathbb A + N \mathbb B.
\end{equation*}
It allows the lapse variable to serve as Lagrange multiplier and constrain the spatial sector while the new shift $n^i$ would be determined by its own (algebraic) equation of motion.

Namely, one can check by direct calculation \cite{RGPMG} that the matrices $\mathbb A$ and $\mathbb B$ are given by
\[
\mathbb A^i_{\phantom{.}k} = \dfrac{1}{\sqrt{1 - n^r \delta_{r s} n^s}}\begin{pmatrix}
	1 & n^r \delta_{r k} \\
	-n^i & -n^i n^r \delta_{r k}
\end{pmatrix}; \quad 
\mathbb B = \begin{pmatrix}
	0 & 0 \\
	0 & \sqrt{(\gamma^{i s} - D^i_{\phantom{.}r} n^r n^l D_{l}^{\phantom{.}s}) \delta_{s k}}
\end{pmatrix};
\]
provided that the transformation of shifts is obtained by solving equation
\begin{equation*}
\left(\sqrt{1 - n^r \delta_{r s} n^s}\right) D^i_{\phantom{.} k} = \sqrt{\left(\gamma^{i s} - D^i_{\phantom{.}r} n^r n^l D_{l}^{\phantom{.}s}\right) \delta_{s k}}
\end{equation*}

The equation of motion for $n^i$ reads
\[
\left(R_i + \dfrac{2 m^2 \sqrt{\gamma} n^l \delta_{l i}}{\sqrt{1 - n^r \delta_{r s} n^s}}\right)\cdot \left(\delta^{i}_{\phantom{.} k} + N \dfrac{\partial}{\partial n^k} (D^i_{\phantom{.} j} n^j) \right) = 0
\]
where the second factor is the Jacobian of the transformation (\ref{Nn}) and cannot be equal to zero, while vanishing of the first factor determines the shift:
\[
n^i = -R_j \delta^{j i} \left[4 m^4 \text{det} \gamma + R_k \delta^{k l} R_l\right]^{-1/2}.
\]
The variation with respect to $N$ gives then:
\begin{equation*}
R^0 + R_i D^i_{\phantom{.} j} n^j + 2 m^2 \sqrt{\gamma} \left[3 - \sqrt{1 - n^r \delta_{r s} n^s} D^k_{\phantom{.} k}\right] = 0
\end{equation*}
which, after substitution of the shifts, is the constraint equation for the spatial sector. Its conservation in time would amount to another constraint of a second class pair. At the final step of the Hamiltonian analysis the value of the lapse would be determined.

Let us note that this proof allows for at least some freedom in choosing the square root $\mathbb X$. Indeed, the $3\times 3$-dimensional square root remains unspecified. And the freedom of changing the sign of the temporal eigenvalue can be realised by simply changing the overall sign of the $\sqrt{\mathbb X}$ together with explicit three-dimensional freedom. However, we see that changing the signs of $\mathbb A$ and $\mathbb B$ does not alter the conclusions more than simply changing the sign of the shift: $n^i = +R_j \delta^{j i} \left[4 m^4 \text{det} \gamma + R_k \delta^{k l} R_l\right]^{-1/2}.$ Therefore, it seems that the proof of Hassan and Rosen has pretty much generic consequences, beyond the standard choice of the square roots.

One can also give independent arguments which look even more generic \cite{me}. It is possible to write down the $\beta_1$ term as $\beta_1\sqrt{\gamma}\left[\mathbb F \right]$ with the constraint, enforced by Lagrange multipliers, that ${\mathbb F}^2=N\mathbb X$. Then one can show that there appears a constraint in the spatial sector \cite{me}. Therefore it might be sensible to consider massive gravity with non-standard square roots. And we must understand the mathematical structures behind them.

\section{The structure of square roots}
\label{Root}

The mathematics of extracting square roots of matrices amounts to solving a simple matrix equation 
\begin{equation}
\label{A2X}
\mathbb A\cdot \mathbb A=\mathbb X
\end{equation}
where $\mathbb X$ is a given $n\times n$ matrix ($g^{-1}f$ in our case), and $\mathbb A$ is an unknown which we denote by $\mathbb A\equiv \sqrt{\mathbb {X}}$.

Note that the procedure of squaring a matrix is respected by similarity transformations in the sense that
\begin{equation}
\label{SXS}
\left({\mathcal S}\mathbb A {\mathcal S}^{-1}\right)^2={\mathcal S} \mathbb A^2{\mathcal  S}^{-1}
\end{equation}
for any non-singular matrix $\mathcal S$. As usual, it is more convenient to start the discussion over the field of complex numbers. As is well known \cite{G}, any matrix can be brought to the Jordan normal form by a similarity transformation. Therefore, it is enough to understand how to take square roots of matrices in the Jordan normal form, for otherwise a suitable similarity transformation can be applied.

It can be shown, see Appendix A and Ref. \cite{G}, that a square root of a Jordan normal form is obtained by taking square roots of separate Jordan blocks\footnote{This is the only possibility if the eigenvalues of different Jordan blocks are not equal to each other. Otherwise there is also an additional continuous freedom in choosing a square root, see below.} which is always possible (recall that we are working with complex numbers) apart from the case of degenerate ($\lambda=0$) Jordan blocks of dimension greater than one. The latter case does not concern us much since our matrices are by definition non-degenerate (see however \cite{GHW1, GHW2}). The details are given in the Appendix A, while here we assume for simplicity that the matrix $\mathbb X$ is diagonalisable, i.e. all its Jordan blocks are $1$-dimensional.

For any diagonalisable matrix $$\mathbb X={\mathcal S}^{-1}\cdot \mathrm{diag}(\lambda_1, \lambda_2, \ldots \lambda_n)\cdot{\mathcal S}$$ we can take
$$\sqrt{\mathbb X}={\mathcal S}^{-1}\cdot \mathrm{diag}\left(\pm\sqrt{\lambda_1}, \pm\sqrt{\lambda_2}, \ldots \pm\sqrt{\lambda_n}\right)\cdot\mathcal S$$
which features discrete freedom which corresponds to $2^n$ elements in $\left(\mathbb Z_2\right)^n$. 

If some eigenvalues are equal then also a larger freedom appears. Indeed, if we have chosen $+\sqrt{\lambda}$ in one place and $-\sqrt{\lambda}$ in another, then a non-trivial similarity transformation in the subspace spanned by corresponding eigenvectors would change the square root $\sqrt{\mathbb X}$ but not the matrix $\mathbb X$ itself, for the latter is proportional to identity in this subspace.

Suppose that the continuous freedom is present. With a given choice of eigenvalues for $\sqrt{\mathbb X}$, the set of different solutions for $\sqrt{\mathbb X}$ naturally has a structure of smooth manifold. Indeed, the group of similarity transformations can be viewed as the adjoint representation of the general linear group $GL(n,\mathbb C)$ given by ${\mathfrak L}_{\mathcal S}:\ {\mathbb X}\longrightarrow {\mathcal S}{\mathbb X}{\mathcal S}^{-1}$. 
The stabiliser subgroup (or centraliser) of a given matrix $\mathbb X$ is, by definition, a subgroup which maps $\mathbb X$ to itself, ${\mathfrak S}_{\mathbb X}\equiv\{{\mathcal S}\in GL(n,\mathbb C):\ {\mathfrak L}_{\mathcal S}\mathbb X=\mathbb X\}$. Obviously, ${\mathfrak S}_{\sqrt{\mathbb X}}\subset{\mathfrak S}_{\mathbb X}$, and the continuous freedom arises when ${\mathfrak S}_{\sqrt{\mathbb X}}\neq{\mathfrak S}_{\mathbb X}$ which is true if and only if equal eigenvalues of $\mathbb X$ were chosen to have unequal square roots. In this case ${\mathfrak S}_{\sqrt{\mathbb X}}$ naturally acts on ${\mathfrak S}_{\mathbb X}$, and the freedom of choosing the square root is parametrised by the homogeneous space \cite{S} ${\mathfrak S}_{\mathbb X}/{\mathfrak S}_{\sqrt{\mathbb X}}$.

\subsection{On conditions of reality}

Of course, we are interested in classifying the real square roots. In this case, one has to bear in mind that a real matrix can have a pair of complex eigenvalues, complex conjugate to each other. As an example, note that a matrix $\begin{pmatrix}
		a & b \\
		-b & a \\
    \end{pmatrix}$ has the normal form which reads $\begin{pmatrix}
		a+ib & 0 \\
		0 & a-ib \\
    \end{pmatrix}$. In this case we need to choose the square roots $\sqrt{a\pm ib}$ to be complex conjugate to each other if we want the inverse similarity transformation to bring it back to real values. It narrows the freedom.
    
Note also that if the matrix $\mathbb X$ has negative eigenvalues, then their square roots are imaginary. However, if they go in pairs, or more generally if there are even numbers of identical Jordan blocks with negative eigenvalues, then the $\sqrt{\lambda}$-s can be chosen in complex conjugate pairs (which of course invokes continuous freedom), and the matrix $\sqrt{\mathbb X}$ would become real after a suitable similarity transformation from what is allowed by the continuous freedom. Otherwise, a real square root does not exist. See also the Ref. \cite{Kocic} for geometrical and physical understanding of reality conditions.

\subsection{The problem with variations}

In the standard approach to massive gravity we make an assumption that the object $\delta \sqrt{\mathbb X} \equiv \sqrt{\mathbb X + \delta \mathbb X} - \sqrt{\mathbb X}$ does exist as an infinitesimal variation of $\sqrt{\mathbb X}$. This assumption can break down with non-standard square roots \cite{NBMG}. This is somewhat unusual. Indeed, usually we can even have a series expansion in powers of $\sqrt{\mathbb X}$, at least around a solution with proportional metrics for which $\mathbb X=c\cdot\mathbb I$ with $c\in\mathbb R$. However, even for proportional backgrounds, if we choose $\sqrt{\mathbb I}\neq\mathbb I$ then a generic $\delta\mathbb X$ does not commute with $\sqrt{\mathbb I}$, and we cannot write the usual expansion for $\sqrt{\mathbb I + \delta \mathbb X}$ in terms of powers of $\delta\mathbb X$.

It makes the standard variations problematic. And actually, in certain cases, the problem is more profound than just a mere technicality. In cases of continuous freedom a smooth variation of $\sqrt{\mathbb X}$ does not exist at all. Let us consider the $2\times 2$ unit matrix $\mathbb I=\left(\begin{matrix} 1 & 0 \\ 0 & 1\end{matrix}\right)$. It has two isolated roots, $\sqrt{\mathbb I}_{++}=\mathbb I=\left(\begin{matrix} 1 & 0 \\ 0 & 1\end{matrix}\right)$ and $\sqrt{\mathbb I}_{--}=-\mathbb I=\left(\begin{matrix} -1 & 0 \\ 0 & -1\end{matrix}\right)$, and also a two-dimensional manifold of roots $\sqrt{\mathbb I}_{ab}=\mathbb I=\left(\begin{matrix} a & b_1 \\ b_2 & -a\end{matrix} \right)$ where $a^2=1-b_1 b_2$. However, even an infinitesimal change of the initial matrix leaves only two points from this manifold\footnote{Introducing new variables $b_{\pm}\equiv\frac{1}{\sqrt2}(b_1\pm b_2)$, we see that the manifold is the surface $a^2+b_{+}^2=1+b_{-}^2$ in three-dimensional space.}. Indeed, for $\mathbb M_{\epsilon}=\left(\begin{matrix} 1 & 0 \\ 0 & 1+\epsilon\end{matrix}\right)$ we have only four isolated roots: $\sqrt{\mathbb M_{\epsilon}}_{++}=\left(\begin{matrix} 1 & 0 \\ 0 & \sqrt{1+\epsilon}\end{matrix}\right)$, \ $\sqrt{\mathbb M_{\epsilon}}_{--}=\left(\begin{matrix} -1 & 0 \\ 0 & -\sqrt{1+\epsilon}\end{matrix}\right)$, \ $\sqrt{\mathbb M_{\epsilon}}_{+-}=\left(\begin{matrix} 1 & 0 \\ 0 & -\sqrt{1+\epsilon}\end{matrix}\right)$, and $\sqrt{\mathbb M_{\epsilon}}_{-+}=\left(\begin{matrix} -1 & 0 \\ 0 & \sqrt{1+\epsilon}\end{matrix}\right)$.

Let us recapitulate the situation. If the freedom is only of discrete type, then it is basically  a technical issue of finding a proper series expansion when $\mathbb X$ and $\delta\mathbb X$ do not commute \cite{Ohta}. For example, the linear correction solves the equation of Sylvester type: $\sqrt{\mathbb X}\cdot\delta\sqrt{\mathbb X}+\delta\sqrt{\mathbb X}\cdot\sqrt{\mathbb X}=\delta\mathbb X$. And the problem is to find $\delta\sqrt{\mathbb X}$ explicitly. It can be done at the expense of using clever field redefinitions \cite{Angi}. 

Continuous freedom presents a much worse problem. Typically in symmetric set-ups, we can have coincident eigenvalues in the background matrix $\mathbb X$, and therefore the freedom of continuous type in $\sqrt{\mathbb X}$. Generic perturbation $\delta\mathbb X$ would split the eigenvalues, and only some discrete subset of possible solutions for $\sqrt{\mathbb X}$ would be smoothly related to those for $\sqrt{\mathbb X + \delta \mathbb X}$. Perturbation theory in terms of matrices is not well defined. Note that the approach with the Sylvester equation works well if and only if the spectra of $\sqrt{\mathbb X}$ and $-\sqrt{\mathbb X}$ have empty intersection \cite{Angi, Sylv}. Obviously, it fails precisely when some equal eigenvalues have different square roots and continuous freedom exists.

Note though that the continuous freedom does not affect the eigenvalues which are the only thing which enters the action. Therefore, it should be better to perform the variations at the level of eigenvalues. We have recently proposed one such method \cite{Dealing17}. In this approach we are not interested in square root matrices, instead we define the polynomials $e_i$ in the action (\ref{TA}) as solutions of algebraic equations in terms of elementary symmetric polynomials of the $g^{-1}f$ or $f^{-1}g$ matrix. Effectively, it eliminates the continuous freedom since the eigenvalues are not changing under the similarity transformations.

\section{New approach to the action principle}
\label{pol}
As we have seen, working in terms of eigenvalues could allow to overcome the problems with perturbation theory which arise when continuous freedom is present. However, explicitly computing eigenvalues for a generic matrix in massive gravity might be a very unpleasant task. In our previous paper \cite{Dealing17} we offered a better method. The action (\ref{TA}) depends only on elementary symmetric polynomials of eigenvalues. And we used the following observation:
\begin{multline}
\label{observation}
\sum_{n=0}^N (-\lambda^2)^{N-n}\cdot e_n(\mathcal M^2)={\rm det}\left({\mathcal M^2}-\lambda^2{\mathbb I}\right)={\rm det}\left(({\mathcal M}-\lambda{\mathbb I})\cdot({\mathcal M}+\lambda{\mathbb I})\right)\\
={\rm det}\left({\mathcal M}-\lambda{\mathbb I}\right)\cdot{\rm det}\left({\mathcal M}+\lambda{\mathbb I}\right)=\left(\sum_{k=0}^N (-\lambda)^{N-k}\cdot e_k(\mathcal M)\right)\cdot\left(\sum_{m=0}^N \lambda^{N-m}\cdot e_m(\mathcal M)\right)
\end{multline}
to find how the symmetric polynomials of $\mathbb X$ and $\sqrt{\mathbb X}$ are related to each other.

From now on we denote for brevity
$$e_i(\mathbb X)\equiv p_i,\quad e_i(\sqrt{\mathbb X})\equiv e_i$$
and get the desired relation from (\ref{observation}) with $\mathcal M=\sqrt{\mathbb X}$:
\begin{equation}
\label{eqep}
\sum\limits_{k+m=2n}(-1)^k e_k e_m=(-1)^n p_n.
\end{equation}
In four dimension we have more explicitly:
\begin{eqnarray}
\label{eq4ep1}
e_1^2 - 2 e_2 & = & p_1,\\
\label{eq4ep2}
e_2^2 - 2 e_1 e_3 + 2e_4 & = & p_2,\\
\label{eq4ep3}
e_3^2- 2 e_2 e_4 & = & p_3,\\
\label{eq4ep4}
e_4^2 & = & p_4.
\end{eqnarray}

The interaction Lagrangian takes the form of $\sum\beta_i e_i$. And we can perform the variation as:
\begin{equation}
\label{varep}
\dfrac{\partial }{\partial g^{\mu \nu}}\left(\sqrt{-g}e_i\right) = \frac{1}{2}\sqrt{-g}\left(-e_i g_{\mu\nu} + 2 \sum_{m} \frac{\partial e_i}{\partial p_m} \frac{\partial p_m}{\partial g^{\mu\nu}}\right)
\end{equation}
where one easily gets
\begin{equation}
\label{pvar}
g^{-1} \dfrac{\partial p_n}{\partial g^{-1}}  =  \sum_{i=1}^n (-1)^{i-1} p_{n-i}\cdot {\mathbb X}^i.
\end{equation}
Indeed, for $n=1$ we trivially see that $g^{\mu\alpha}\frac{\partial}{\partial g^{\alpha\nu}}\left[g^{-1}\eta\right]={\mathbb X}^{\mu}_{\nu}$ and more generally $g^{-1}\frac{\partial}{\partial g^{-1}}\left[\mathbb X^n\right]=n {\mathbb X}^n$, and then go by induction using $p_n=\frac{1}{n}\sum\limits_{i=1}^n (-1)^{i-1}[\mathbb X^i]\cdot p_{n-i}$.

In any case when the standard approach in terms of matrices works well, it is guaranteed that these equations of motion coincide with the usual ones since we are doing literally the same variation though without having the matrix elements explicitly. Let us illustrate how it works in the simplest case of two dimensions where equations (\ref{eqep}) take the form of
\begin{eqnarray*}
e_1^2 - 2 e_2 & = & p_1,\\
e_2^2 & = & p_2.
\end{eqnarray*}
and therefore
\begin{equation*}
\frac{\partial p}{\partial e}=2\left(\begin{matrix}e_1 & -1 \\ 0 & e_2 \end{matrix}\right); \qquad \frac{\partial e}{\partial p}=\frac{1}{2e_1e_2}\left(\begin{matrix}e_2 & 1 \\ 0 & e_1 \end{matrix}\right).
\end{equation*}
After a simple calculation we obtain from (\ref{varep}) and (\ref{pvar})
\[
\dfrac{\partial }{\partial g^{\mu \nu}}\left(\sqrt{-g}e_1\right) = \frac{1}{2}\sqrt{-g}\left(\frac{e_2+p_1}{e_1e_2}\eta_{\mu\nu}-\frac{1}{e_1 e_2}\eta_{\mu\alpha}g^{\alpha\beta}\eta{\beta\nu}-e_1g_{\mu\nu}\right).
\]
Multiplying this relation by $g^{-1}$, we need to prove that ${\mathcal M}^2=\mathbb X$ for 
$$\mathcal M=\frac{e_2+p_1}{e_1e_2}\mathbb X-\frac{1}{e_1 e_2}\mathbb X^2=\frac{1}{e_1e_2}\left(e_2 \mathbb X+e_2^2 \mathbb I \right)$$
where we have used the Cayley-Hamilton theorem (see Appendix A.5), $\mathbb X^2-p_1\mathbb X+p_2\mathbb I=\mathbb O$, for the matrix $\mathbb X$. It can be done by direct calculation employing again the Cayley-Hamilton theorem.

In four dimensions the things become rather bulky, and we have
\[
\frac{\partial e_i}{\partial p_m}=\left( \frac{\partial p}{\partial e} \right)^{-1}=\dfrac{1}{{\rm det}\left(\frac{\partial p}{\partial e}\right)}
\begin{pmatrix}
e_4 (e_2 e_3 - e_1 e_4) & e_3 e_4 & e_1 e_4 & e_1 e_2 - e_3 \\
e_3^2 e_4 & e_1 e_3 e_4 & e_1^2 e_4 & e_1 (e_1 e_2 - e_3) \\
e_3 e_4^2 & e_1 e_4^2 & e_1 e_2 - e_3 e_4 & -e_2 e_3 + e_1 (e_2^2 - e_4) \\
0 & 0 & 0 & e_1 e_2 e_3 - e_3^2 - e_1^2 e_4
\end{pmatrix}
\]
where
\begin{equation}
\label{det4}
{\rm det}\left(\frac{\partial p}{\partial e}\right)_{4\times 4}=2 e_4 (e_1 e_2 e_3 -  e_3^2 - e_1^2 e_4),
\end{equation}
and for the $\beta_1$ term we have to check again that ${\mathcal M}^2=\mathbb X$ with
\[
\mathcal M = \dfrac{e_1^2 e_3 \mathbb X + e_1^3 \mathbb X^2 - e_1 (e_2 e_4 \mathbb I + (e_2^2 + e_4) \mathbb X + 2 e_2 \mathbb X^2 + \mathbb X^3) + e_3 (e_4 \mathbb I + e_2 \mathbb X + \mathbb X^2)}{-e_1 e_2 e_3 + e_3^2 + e_1^2 e_4}.
\]
where we have used, and need to further use in  ${\mathcal M}^2$, the Cayley–Hamilton theorem which in four dimensions reads $\mathbb X^4 - p_1 \mathbb X^3 + p_2 \mathbb X^2 - p_3 \mathbb X + p_4 \mathbb I=\mathbb O$. After we have got an expression for $\sqrt{\mathbb X}$, all other terms can be considered straightforwardly.

Obviously, it can be done in any dimension, and works well as long as ${\rm det}\left(\frac{\partial p}{\partial e}\right)\neq 0$. However, we will not go into this combinatorics, and we only write down the determinant in the three dimensional case
\begin{equation}
\label{det3}
{\rm det}\left(\frac{\partial p}{\partial e}\right)_{3\times 3}=2 e_3 (e_1 e_2  -  e_3)
\end{equation}
for future use.

\subsection{Perturbation theory}

Obviously, it is not always possible to have an exact solution, and one often uses perturbation theory. In particular, for the weak gravity limit the background solution corresponds to $\mathbb X=\mathbb I$ and values of polynomials $p_1=4,\quad p_2=6,\quad p_3=4,\quad p_4=1$. With the trivial choice of the square root $\sqrt{\mathbb X}=\mathbb I$ we have $e_i=p_i$, and one can check that ${\rm det}\left(\frac{\partial p}{\partial e}\right)\neq 0$.

For perturbative calculations around Minkowski space in massive gravity it is convenient to use the symmetry of massive gravity (\ref{gfsymmetry}) and work with $\eta^{-1}g\equiv\mathbb I+\mathcal H$ matrix instead of $g^{-1}\eta$ since $\sqrt{-g}e_i\left(\sqrt{g^{-1}\eta}\right)=e_{4-i}\left(\sqrt{\eta^{-1}g}\right)$, and having worked out a symmetric polynomial one does not need to multiply its perturbative expansion by expansion for $\sqrt{-g}$.

In the previous paper \cite{Dealing17}, we have shown how it goes, and reproduced the standard result that Minkowski space is a vacuum solution if and only if 
$$\beta_0=-3\beta_1-3\beta_2-\beta_3,$$
and the quadratic action is then
\begin{multline*}
\mathcal L_{\rm int}=\sum_{n=0}^3 \beta_n\cdot e_{4-n}\left(\sqrt{\eta^{-1}g}\right)=\frac{1}{8}\left(\beta_1+2\beta_2+\beta_3\right)\cdot\left(h^{\mu\nu}h_{\mu\nu}-(h^{\mu}_{\mu})^2\right)\\
=-\frac14\left(\beta_1+2\beta_2+\beta_3\right)\cdot e_2\left(\mathcal H\right)
\end{multline*}
which is the Fierz-Pauli term.

\section{Perturbation theory around unusual roots}
\label{Mink}

Now let us consider perturbation theory around unusual roots. Note that in those cases when odd numbers of eigenvalues are taken with reversed signs, we need to be more accurate about the symmetry transformation (\ref{gfsymmetry}). Since any such choice entails ${\rm det}\left(\sqrt{\eta^{-1}g}\right)<0$, then we need to take $\beta_i\leftrightarrow -\beta_{D-i}$  instead of $\beta_i\leftrightarrow \beta_{D-i}$ if $\sqrt{-g}$ is kept positive. This is not a big deal, but unfortunately it is not the end of the story.

Unusual roots are precisely the cases when ${\rm det}\left(\frac{\partial p}{\partial e}\right)= 0$. And perturbation theory ceases to be analytic and nice. Let us first look at a toy example of $2D$. Unusual square roots are given by 
$$\sqrt{\mathbb I}=\left(\begin{matrix}-1 & 0\\ 0 & 1 \end{matrix}\right)$$ 
and all its similarity transformations. In other words, we have $e_1=0, \quad e_2=-1$ in the background. Then we have $p_1=2+e_1(\mathcal H)$ and $p_2=1+e_1(\mathcal H)+e_2(\mathcal H)$ with perturbation $\mathcal H\equiv \eta^{-1}\cdot\delta g$. Equations for polynomials with our background choice can be solved exactly as 
\begin{eqnarray*}
e_2  =  -\sqrt{p_2} & = & -\sqrt{1+e_1(\mathcal H)+e_2(\mathcal H)},\\
e_1  =  \pm \sqrt{p_1+2e_2} & = & \pm\sqrt{2+e_1(\mathcal H)-2\sqrt{1+e_1(\mathcal H)+e_2(\mathcal H)}}.
\end{eqnarray*}
We see that there are two solutions for $e_1$. Moreover, at the lowest order we have
$$e_1=\pm\frac12\sqrt{\left(e_1(\mathcal H)\right)^2-4e_2(\mathcal H)}+{\mathcal O}\left(\mathcal H^2\right)$$
which scales linearly with $\mathcal H$ but not analytic.

The reason is simple. For a generic perturbation $\mathcal H$ two eigenvalues are different, and the sign of $e_1$ depends on which of them (the smaller or the larger) is taken with the minus sign.

In the interaction potential we have then ${\mathcal L}_{int}=-\beta_0 e_2-\beta_1 e_1$ which has at the linear level $e_2=-1-\frac12 {\mathcal H}^{\mu}_{\mu}+{\mathcal O}\left({\mathcal H}^2\right)$ and $e_1=\pm\sqrt{\frac12 {\mathcal H}^{\mu\nu}{\mathcal H}_{\mu\nu}-\frac14\left({\mathcal H}^{\mu}_{\mu}\right)^2}\cdot\left(1+{\mathcal O}({\mathcal H})\right)$. We see that vanishing of the linear variation with arbitrary ${\mathcal H}$ around this solution requires two conditions, for analytic and non-analytic parts separately. In $2D$ it is too restrictive, and sets both $\beta$-s to zero.

Obviously, non-analytic features should be seen in more complicated cases, too. The number of solutions is given by the number of possible ways to choose the distribution of signs among the square roots of eigenvalues. We will now check it explicitly in three and four dimensions.

\subsection{The 3D case}

Let us now turn to three dimensions. Since we are interested in Minkowski background, we need to discuss square roots of three-dimensional unit matrix. In terms of eigenvalues, we choose the signs of three square roots of unity. Barring the overall sign of the potential term, there are two distinct options: either all three square roots have equal signs which is the standard solution without continuous freedom, or we have one eigenvalue with a different sign.
We see that in three dimensions it suffices to study the following unusual square root of $\mathbb I$:
\begin{equation}
\label{sqrt3}
\left.\sqrt{\eta^{-1}g}\ \right|_{g=\eta}=\sqrt{\mathbb I}=
\begin{pmatrix}
-1 & 0 & 0 \\
0 & 1 & 0 \\
0 & 0 & 1
\end{pmatrix}
\end{equation}
and all its similarity transformations.

We have the interaction term
$$\mathcal L_{\rm int}=\sqrt{-g} \sum_{n=0}^2 \beta_n\cdot e_n=-\sum_{n=0}^2 \beta_n\cdot e_{3-n}$$
where the prefactor of $m^2M_{Pl}^2$ is omitted (or absorbed into the $\beta$-s). Also the $\beta_3$ term has been omitted since it does not depend on $g$. The difference from the standard approach is that we now treat the quantities $e_i$ as solutions of algebraic equations (\ref{eqep}) with no regard to square root matrices which effectively eliminates the continuous freedom of similarity transformations. Equations (\ref{eqep}) take the form
\begin{eqnarray}
\label{eq3ep1}
e_1^2- 2 e_2 & = & p_1,\\
\label{eq3ep2}
e_2^2- 2 e_1e_3 & = & p_2,\\
e_3^2 & = & p_3.
\label{eq3ep3}
\end{eqnarray}
with $p_i\equiv e_i\left(\eta^{-1}g\right)=e_i\left(\mathbb I + \mathcal H\right)$:
\begin{eqnarray*}
p_1 & = & 3 + h_1,\\
p_2 & = & 3 + 2h_1 + h_2,\\
p_3 & = & 1 + h_1+ h_2 + h_3.
\end{eqnarray*}
where $h_i\equiv e_i(\mathcal H)$ are elementary symmetric polynomials of the matrix ${\mathcal H}^{\mu}_{\nu}=\eta^{\mu\alpha}\cdot\delta g_{\alpha\nu}$.

We now build the perturbation theory in the form of 
\begin{equation}
\label{series}
e_k=\sum_{i\geq 0} e^{(i)}_k
\end{equation}
with $e^{(i)}_k$ being the $i$-th order (which scales as the $i$-th power of $\mathcal H$) correction to the $k$-th polynomial. The $0$-th orders are by definition the background values: $e^{(0)}_1=1$,\quad $e^{(0)}_2=-1$,\quad $e^{(0)}_3=-1$ for which the determinant in the formula (\ref{det3}) is obviously equal to zero. Nevertheless, the third polynomial can be directly solved for according to (\ref{eq3ep3}):
\begin{multline}
\label{e3}
e_3=-1-\frac12 h_1-\frac12\left(h_2-\frac14 h_1^2\right)-\frac12\left(h_3-\frac12 h_1h_2+\frac18 h_1^3\right)\\
+\frac14 \left(h_1h_3+\frac12 h_2^2-\frac34 h_1^2 h_2+\frac{5}{32}h_1^4\right)+{\mathcal O}\left({\mathcal H}^5\right)
\end{multline}

Now we substitute the expansions (\ref{series}) and (\ref{e3}) into the system of equations (\ref{eq3ep1}), (\ref{eq3ep2}) and keep only terms of the first order:
\begin{eqnarray*}
2 e^{(1)}_1 - 2 e^{(1)}_2 &=& h_1, \\
2 e^{(1)}_1 - 2 e^{(1)}_2 + 2\cdot \frac12 h_1 &=& 2h_1.
\end{eqnarray*}
The system is obviously degenerate. We have
\begin{equation}
\label{x21}
e^{(1)}_2=e^{(1)}_1-\frac12 h_1
\end{equation}
with one first order variable undetermined at the linear level.

Taking this (\ref{x21}) into account, we get again two coincident equations (\ref{eq3ep1}), (\ref{eq3ep2}) at quadratic order:
\begin{equation*}
\left(e^{(1)}_1\right)^2 + 2 e^{(2)}_1 - 2 e^{(2)}_2 = 0
\end{equation*}
which can be solved as
\begin{equation}
\label{x22}
e^{(2)}_2=e^{(2)}_1+\frac12 \left(e^{(1)}_1\right)^2
\end{equation}
now leaving one second order variable undetermined.

Let us now turn to the third order. The first equation (\ref{eq3ep1}) acquires the form
\begin{equation*}
e^{(3)}_2 = e^{(3)}_1 + e^{(1)}_1 e^{(2)}_1,
\end{equation*}
and, after taking it into account, the second one (\ref{eq3ep2}) finally allows to know the first order completely:
\begin{equation}
\label{cubicguy}
\left(e^{(1)}_1\right)^3-\frac12 h_1 \left(e^{(1)}_1\right)^2+\left(h_2-\frac14 h_1^2\right)e^{(1)}_1+h_3-\frac12 h_1 h_2+\frac18 h_1^3=0.
\end{equation}
This is a generic cubic equations (the  three coefficients depend on three parameters $h_i$). Its three roots correspond to freedom of choosing which one of three eigenvalues of perturbed matrix $\mathbb X$ would go with the reversed sign.

It is worth to perform a consistency check here. Let $\lambda_i$, $i=1,2,3$ be eigenvalues of $\mathcal H$. Then we have $h_1=\sum\limits_i \lambda_i$,\quad $h_2=\sum\limits_{i<j} \lambda_i\lambda_j$,\quad $h_3=\sum\limits_{i<j<k} \lambda_i\lambda_j\lambda_k=\lambda_1\lambda_2\lambda_3$. One can easily check by Viet theorem that the three solutions of the cubic equation (\ref{cubicguy}) are $\frac12\left(-\lambda_1+\lambda_2+\lambda_3\right)$,\quad $\frac12\left(-\lambda_2+\lambda_3+\lambda_1\right)$, and $\frac12\left(-\lambda_3+\lambda_1+\lambda_2\right)$. These are indeed the three possible first order contributions to $e_1$ with eigenvalues $\pm\sqrt{1+\lambda_i}$ and our choice of the square root.

After we have reproduced the full freedom of choosing the eigenvalues, the system (\ref{eq3ep1}), (\ref{eq3ep2}) becomes non-degenerate giving two new pieces of information at each order. In particular, at the fourth order the first equation (\ref{eq3ep1}) gives
\begin{equation}
\label{x24}
e^{(4)}_2=e^{(4)}_1+e^{(1)}_1 e^{(3)}_1+\frac12 \left(e^{(2)}_1\right)^2.
\end{equation}
At the same time, the second equation (\ref{eq3ep2}) contains $e^{(2)}_1$ both quadratically and linearly. However, after taking the first equation (\ref{x24}) into account, one gets a linear equation for $e^{(2)}_1$ which can be unequivocally solved. Explicit expressions are rather cumbersome, and we don't go into that in this paper\footnote{To avoid confusion, let us mention that there would be a denominator which can take the zero value. However, playing with $\lambda$-s one can check that it happens only when the enumerator is zero, too.}.

The first order contribution to the interaction Lagrangian is:
\[
\delta^{(1)}\mathcal L_{\rm int} = \frac12\beta_0 h_1-\beta_1 \left(e^{(1)}_1-\frac12 h_1\right)-\beta_2 e^{(1)}_1.
\]
If Minkowski space is to be a solution of equations of motion, then we need to impose  $\delta^{(1)}\mathcal L_{\rm int}=0$. Since $e^{(1)}_1$ is generically irrational for rational values of $h_i$, we need the two parts, with $h_1$ and with $e^{(1)}_1$, to vanish independently: 
\begin{equation}
\label{beta3D}
\beta_0=-\beta_1,\quad \beta_2=-\beta_1.
\end{equation}

It is interesting to note that one can formally plug the Minkowski space with our choice (\ref{sqrt3}) of $\sqrt{g^{-1}\eta}$ into the naive equation of motion (\ref{eqg}),
$$0=\sum_{i=0}^2 (-1)^i\beta_i \mathbb Y_{(i)}=\begin{pmatrix}
\beta_0+2\beta_1+\beta_2 & 0 & 0 \\
0 & \beta_0-\beta_2 & 0 \\
0 & 0 & \beta_0-\beta_2
\end{pmatrix},$$
and deduce the same condition (\ref{beta3D}) for $\beta$-s.

Now we look at the second order Lagrangian for perturbations:
\begin{multline*}
\delta^{(2)}\mathcal L_{\rm int} = \frac12\beta_0 \left(h_2-\frac14 h_1^2\right)-\beta_1 \left(e^{(2)}_1+\frac12 \left(e^{(1)}_1\right)^2\right)-\beta_2 e^{(2)}_1\\
=-\frac{\beta_1}{2}\cdot\left(\left(e^{(1)}_1\right)^2+h_2-\frac14 h_1^2\right).
\end{multline*}
We see that it contains the root $e^{(1)}_1$ of the cubic equation (\ref{cubicguy}), and therefore is not analytic in $\mathcal H$. If we enumerate the eigenvalues of $\mathcal H$ such that $e^{(1)}_1 = \dfrac12 (-\lambda_1 + \lambda_2 + \lambda_3)$, we get:
\[
\delta^{(2)}\mathcal L_{\rm int}  = -\frac{\beta_1}{2}\cdot \lambda_2 \lambda_3.
\]
The three branches of solution correspond to three possible choices of two out of three eigenvalues for putting them into $\delta^{(2)}\mathcal L_{\rm int} $.

\subsection{Square roots in the physical dimension}

In four dimensions the system of equations (\ref{eqep}) would be slightly more complicated:
\begin{eqnarray}
\label{eq41ep1}
e_1^2- 2 e_2 & = & 4+h_1,\\
\label{eq41ep2}
e_2^2- 2 e_1e_3 + 2 e_4 & = & 6+3h_1+h_2,\\
\label{eq41ep3}
e_3^2- 2 e_2e_4 & = & 4+3h_1+2h_2+h_3,\\
\label{eq41ep4}
e_4^2 & = & 1+h_1+h_2+h_3+h_4.
\end{eqnarray}
Now we basically have two possible choices of unusual roots $\sqrt{\mathbb I}$: with one reversed-sign eigenvalue or with two of them (and again with continuous freedom of similarity transformations which can be disregarded when working directly with eigenvalues).

Let us start with the first option
\begin{equation}
\label{sqrt4one}
\left.\sqrt{\eta^{-1}g}\ \right|_{g=\eta}=\sqrt{\mathbb I}=
\begin{pmatrix}
-1 & 0 & 0 & 0 \\
0 & 1 & 0 & 0 \\
0 & 0 & 1 & 0\\
0 & 0 & 0 & 1
\end{pmatrix}
\end{equation}
which goes pretty much analogous to the three dimensional case. The background values for this choice are $e^{(0)}_1=2$,\quad $e^{(0)}_2=0$,\quad $e^{(0)}_3=-2$,\quad $e^{(0)}_4=-1$, and one easily checks that the Jacobian (\ref{det4}) of transformation from $e_i$ to $p_i$ vanishes.

As before, the last polynomial (\ref{eq41ep4}) is easy to find:
\begin{multline*}
e_4=-1-\frac12 h_1-\frac12\left(h_2-\frac14 h_1^2\right)-\frac12\left(h_3-\frac12 h_1h_2+\frac18 h_1^3\right)\\
-\frac12 \left(h_4-\frac12 h_1h_3-\frac14 h_2^2+\frac38 h_1^2 h_2-\frac{5}{64}h_1^4\right)+{\mathcal O}\left({\mathcal H}^5\right).
\end{multline*}

The remaining equations (\ref{eq41ep1}) - (\ref{eq41ep3}) give at the first order:
\begin{eqnarray*}
4 e^{(1)}_1 - 2 e^{(1)}_2 &=& h_1, \\
4 e^{(1)}_1 - 4 e^{(1)}_3 - h_1 &=& 3h_1,\\
- 4 e^{(1)}_3  + 2 e^{(1)}_2  &=& 3h_1.
\end{eqnarray*}
The system is degenerate, and we can only find that
\begin{eqnarray*}
e^{(1)}_2 &=& 2e^{(1)}_1 -\frac12 h_1, \\
e^{(1)}_3 &=& e^{(1)}_1 -h_1
\end{eqnarray*}
leaving one variable undetermined.

At the second order we get a similar degeneracy, and find
\begin{eqnarray*}
e^{(2)}_2 &=& 2e^{(2)}_1 +\frac12 \left(e^{(1)}_1\right)^2, \\
e^{(2)}_3 &=& e^{(2)}_1 +\frac12 \left(e^{(1)}_1\right)^2-\frac12\left(h_2-\frac14 h_1^2\right).
\end{eqnarray*}

The degeneracy gets broken at the fourth order level producing a generic (in terms of independent parameters) quartic equation for $e^{(1)}_1$.

Let us now have a look at the interaction Lagrangian:
$$\mathcal L_{\rm int}=\sqrt{-g} \sum_{n=0}^3 \beta_n\cdot e_n\left(\sqrt{g^{-1}\eta}\right)=-\sum_{n=0}^3 \beta_n\cdot e_{4-n}\left(\sqrt{\eta^{-1}g}\right)$$
the first order contribution to which reads
\[
\delta^{(1)}\mathcal L_{\rm int} = \frac12\beta_0 h_1-\beta_1 \left(e^{(1)}_1 -h_1\right)-\beta_2 \left(2e^{(1)}_1 -\frac12 h_1\right)-\beta_3 e^{(1)}_1.
\]
Again, the Minkowski space is a solution when  $\delta^{(1)}\mathcal L_{\rm int}=0$. And we require the two parts, with $h_1$ and $e^{(1)}_1$, to vanish independently: 
\begin{equation}
\label{beta4one}
\beta_0 + 2 \beta_1 + \beta_2 = 0,\quad \beta_1 + 2 \beta_2 + \beta_3 = 0.
\end{equation}

As in $3D$, one can formally plug the Minkowski space with (\ref{sqrt4one}) into the naive equations of motion $\sum\limits_{i=0}^3 (-1)^i\beta_i \mathbb Y_{(i)}=0$:
$$\begin{pmatrix}
\beta_0+3\beta_1+3\beta_2+\beta_3 & 0 & 0 & 0 \\
0 & \beta_0+\beta_1-\beta_2-\beta_3 & 0 & 0 \\
0 & 0 & \beta_0+\beta_1-\beta_2-\beta_3 & 0\\
0 & 0 & 0 & \beta_0+\beta_1-\beta_2-\beta_3
\end{pmatrix}=0$$
and deduce the same condition (\ref{beta4one}) for $\beta$-s.

Now we look at the second order Lagrangian for perturbations:
\begin{multline*}
\delta^{(2)}\mathcal L_{\rm int} = \frac12\beta_0 \left(h_2-\frac14 h_1^2\right)-\beta_1 \left(e^{(2)}_1 +\frac12 \left(e^{(1)}_1\right)^2-\frac12\left(h_2-\frac14 h_1^2\right)\right)\\
-\beta_2\left(2e^{(2)}_1 +\frac12 \left(e^{(1)}_1\right)^2\right
)-\beta_3 e^{(2)}_1\\
=-\frac{\beta_1+\beta_2}{2}\cdot\left(\left(e^{(1)}_1\right)^2+h_2-\frac14 h_1^2\right).
\end{multline*}
If we enumerate the eigenvalues of $\mathcal H$ such that $e^{(1)}_1 = \dfrac12 (-\lambda_1 + \lambda_2 + \lambda_3+\lambda_4)$, then we get:
\[
\delta^{(2)}\mathcal L_{\rm int}  = -\frac{\beta_1+\beta_2}{2}\cdot\left(\lambda_2 \lambda_3 + \lambda_1 \lambda_2 + \lambda_1 \lambda_3\right).
\]

The structure is basically the same as for the Fierz-Pauli term, though with only a part of the full set of eigenvalues. We remind to the reader that for such square roots the usual perturbation theory in terms of matrices is not well defined unless the perturbation respects separation into two subspaces (of positive and negative eigenvalues). In our case, of this and the previous subsections, the latter occurs if $h_{1i}=0$ for $i\geq 2$. Then the structure of quadratic Lagrangian would be $\sum\limits_{i,j\geq 2}h_{ij}h^{ij}-\left(\sum\limits_{i\geq 2}h^i_i\right)^2$. Otherwise the perturbation theory is not analytic and has three (respectively four, or $D$) branches according to solutions of the cubic (respectively quartic, or power $D$) equation.

An interesting fact is that the background equation of motion is correctly reproduced by naively applying formula (\ref{eqg}) which is not even well defined in such cases. Apparently, it has to do with the fact that $\mathbb X\ \propto\ \mathbb I$ can be treated as a limiting case of a cosmological matrix ${\rm diag}\left(N(t),a(t), \ldots, a(t)\right)$ for which the perturbations are well defined. Note though that this limit is singular in the sense that it features appearance of non-analyticity in perturbations, and is not defined at all in the language of matrices  (compare to \cite{NBMG} where they also mention non-analyticity in the case of non-zero "new shift" variable which is probably the first indication in the literature towards the phenomena we have thoroughly studied in the present paper).

\subsection{4D: the second case}

Another option available in four dimensions can be represented by the root
\begin{equation}
\label{sqrt4two}
\left.\sqrt{\eta^{-1}g}\ \right|_{g=\eta}=\sqrt{\mathbb I}=
\begin{pmatrix}
-1 & 0 & 0 & 0 \\
0 & -1 & 0 & 0 \\
0 & 0 & 1 & 0\\
0 & 0 & 0 & 1
\end{pmatrix}
\end{equation}
for which the background values are $e^{(0)}_1=0$,\quad $e^{(0)}_2=-2$,\quad $e^{(0)}_3=0$,\quad $e^{(0)}_4=1$; they correspond to vanishing determinant in equation (\ref{det4}). And the last polynomial (\ref{eq4ep4}) is
\begin{multline*}
e_4=1+\frac12 h_1+\frac12\left(h_2-\frac14 h_1^2\right)+\frac12\left(h_3-\frac12 h_1h_2+\frac18 h_1^3\right)\\
+\frac12 \left(h_4-\frac12 h_1h_3-\frac14 h_2^2+\frac38 h_1^2 h_2-\frac{5}{64}h_1^4\right)+{\mathcal O}\left({\mathcal H}^5\right).
\end{multline*}

The first order equations (\ref{eq4ep1}) - (\ref{eq4ep3}) are particularly simple and degenerate:
\begin{eqnarray*}
- 2 e^{(1)}_2 &=& h_1, \\
- 4 e^{(1)}_2 + h_1 &=& 3h_1,\\
- 2 e^{(1)}_2  + 2 h_1  &=& 3h_1
\end{eqnarray*}
which only give
\begin{equation*}
e^{(1)}_2 =  -\frac12 h_1
\end{equation*}
leaving two other variables undetermined.

At the second order the equations (\ref{eq4ep1}) - (\ref{eq4ep3}) are
\begin{eqnarray*}
\left(e^{(1)}_1\right)^2 - 2e^{(2)}_2 &=& 0, \\
-4e^{(2)}_2 -2e^{(1)}_1 e^{(1)}_3 &=& 0, \\
\left(e^{(1)}_3\right)^2 -2e^{(2)}_2 &=& 0,
\end{eqnarray*}
and we deduce
\begin{eqnarray*}
e^{(2)}_2 &=& \frac12 \left(e^{(1)}_1\right)^2, \\
e^{(1)}_3  &=& -e^{(1)}_1.
\end{eqnarray*}

At the third order, the first equation (\ref{eq4ep1}) gives
\begin{equation*}
e^{(2)}_3 =   e^{(1)}_1 e^{(2)}_1,
\end{equation*}
and after that the second  (\ref{eq4ep2}) and the third (\ref{eq4ep3}) equations are degenerate with each other and give an interesting result:
\begin{equation}
\label{x12and32}
2e^{(1)}_1 \left(e^{(2)}_1+e^{(2)}_3\right)=h_3-\frac12 h_1 h_2+\frac18 h_1^3-\frac12 h_1 \left(e^{(1)}_1\right)^2.
\end{equation}

At this point one might start worrying about the self-consistency. Wouldn't the perturbation theory break down at the second order when $e^{(1)}_1=0$? Indeed, given some eigenvalues $\lambda_i$, there should be six values of $e^{(1)}_1$ of the form $\frac12\left(\lambda_1+\lambda_2-\lambda_3-\lambda_4\right)$. Can a third order symmetric polynomial $h_3-\frac12 h_1 h_2+\frac18 h_1^3$ be divisible by all six of them? Actually, it can since those six values come in three pairs of opposite sign numbers. And one can easily check that
$$h_3-\frac12 h_1 h_2+\frac18 h_1^3=\frac18\left(\lambda_1+\lambda_2-\lambda_3-\lambda_4\right)\left(\lambda_1+\lambda_3-\lambda_2-\lambda_4\right)\left(\lambda_1+\lambda_4-\lambda_2-\lambda_3\right)$$
which gives a finite value for
\begin{equation}
\label{laone}
e^{(2)}_1+e^{(2)}_3=\frac12\left(\lambda_3\lambda_4-\lambda_1\lambda_2\right).
\end{equation}

Then, at the fourth order, the system of equations (\ref{eq4ep1}) - (\ref{eq4ep3}) becomes non-degenerate, and the first equation (\ref{eq4ep1}) reads
\begin{equation}
\label{x24bis}
e^{(4)}_2 =   e^{(1)}_1 e^{(3)}_1+\frac12 \left(e^{(2)}_1\right)^2.
\end{equation}
Two other equations are very cumbersome. However, subtracting the second (\ref{eq4ep2}) from the third  (\ref{eq4ep3}), and using the first one (\ref{x24bis}), we get a very nice equation which fully determines the first order of perturbations:
\begin{equation}
\label{quarticguy}
h_4-\frac14\left(\left(e^{(2)}_1\right)^2+h_2-\frac14 h_1^2\right)^2+\left(e^{(2)}_1+e^{(2)}_3\right)^2=0.
\end{equation}
Together with the result (\ref{x12and32}) for $e^{(2)}_1+e^{(2)}_3$, it amounts to cubic equation for $\left(e^{(1)}_1\right)^2$ precisely as it should be.

Moreover, taking again $e^{(1)}_1=\frac12\left(\lambda_1+\lambda_2-\lambda_3-\lambda_4\right)$ we easily see that
\begin{equation}
\label{latwo}
\frac12\left(\left(e^{(2)}_1\right)^2+h_2-\frac14 h_1^2\right)=\frac12\left(\lambda_3\lambda_4+\lambda_1\lambda_2\right)
\end{equation}
and, given also (\ref{laone}), those values of $e^{(1)}_1$ are indeed the solutions of (\ref{quarticguy}).

Let us now have a look at the interaction Lagrangian:
$$\mathcal L_{\rm int}=\sqrt{-g} \sum_{n=0}^3 \beta_n\cdot e_n\left(\sqrt{g^{-1}\eta}\right)=\sum_{n=0}^3 \beta_n\cdot e_{4-n}\left(\sqrt{\eta^{-1}g}\right)$$
with the first order expression being
\[
\delta^{(1)}\mathcal L_{\rm int} = \frac12\beta_0 h_1-\beta_1 e^{(1)}_1 -\frac12\beta_2 h_1+\beta_3 e^{(1)}_1.
\]
We look for Minkowski space to be a solution:  $\delta^{(1)}\mathcal L_{\rm int}=0$, and demand 
\begin{equation}
\label{beta4two}
\beta_0 = \beta_2,\quad \beta_1 = \beta_3.
\end{equation}

Probably, at this point the reader would not be much surprised if we say that the naive equation of motion $\sum\limits_{i=0}^3 (-1)^i\beta_i \mathbb Y_{(i)}=0$ formally gives
$$\begin{pmatrix}
\beta_0+\beta_1-\beta_2-\beta_3 & 0 & 0 & 0 \\
0 & \beta_0+\beta_1-\beta_2-\beta_3 & 0 & 0 \\
0 & 0 & \beta_0-\beta_1-\beta_2+\beta_3 & 0\\
0 & 0 & 0 & \beta_0-\beta_1-\beta_2+\beta_3
\end{pmatrix}=0$$
which entails the same condition (\ref{beta4two}) for $\beta$-s.

The second order Lagrangian for perturbations is
\begin{multline*}
\delta^{(2)}\mathcal L_{\rm int} = \frac12\beta_0 \left(h_2-\frac14 h_1^2\right)+\beta_1 e^{(2)}_3
+\frac12 \beta_2\left(e^{(1)}_1\right)^2+\beta_3 e^{(2)}_1\\
=\frac{\beta_1}{2}\cdot\left(\frac{h_3-\frac12 h_1 h_2+\frac18 h_1^3}{e^{(1)}_1}-\frac12 h_1 e^{(1)}_1\right)+\frac{\beta_2}{2}\cdot\left(\left(e^{(1)}_1\right)^2+h_2-\frac14 h_1^2\right)
\end{multline*}
which, given the identifications (\ref{laone}) and (\ref{latwo}) above, is equivalent to
\[
\delta^{(2)}\mathcal L_{\rm int} = \dfrac{\beta_2 + \beta_1}{2}\cdot \lambda_3 \lambda_4 + \dfrac{\beta_2 - \beta_1}{2}\cdot \lambda_1 \lambda_2. 
\]

The structure is basically the same as for the Fierz-Pauli term, though with two separate parts of the full set of eigenvalues. As we have already mentioned before, the usual perturbation theory in terms of matrices is not well defined unless the perturbation respects separation into two subspaces (of positive and negative eigenvalues). In our case the latter occurs if $h_{ij}=0$ for $i\leq 2$ and $j\geq 3$ or vice versa. Then the structure of quadratic Lagrangian has two Fierz-Pauli-like terms $\sum\limits_{i,j\leq 2}h_{ij}h^{ij}-\left(\sum\limits_{i\leq 2}h^i_i\right)^2$ and$\sum\limits_{i,j\geq 3}h_{ij}h^{ij}-\left(\sum\limits_{i\geq 3}h^i_i\right)^2$. Otherwise the perturbation theory is not analytic. The six branches of it correspond to different separations of four eigenvalues into two groups of two.

Note that this kind of unusual root can also be related to bimetric cosmology. However, making it regular would require at least an axially symmetric Bianchi I spacetime. In bimetric Friedmann cosmology given by $$g=\text{diag}\{1, a^2(t), a^2(t), a^2(t)\},\quad f=\text{diag} \{W^2(t), Y^2(t), Y^2(t), Y^2(t)\}$$ with  $\mathbb X=g^{-1}f=\{W^2, Y^2/a^2, Y^2/a^2, Y^2/a^2\}$, the classical equations of motion are well defined only for square roots of $\sqrt{g^{-1}f} =\pm \{\pm W, Y/a, Y/a, Y/a\}$ type since they do not make equal eigenvalues different. But we must be careful if choosing another square root: $\sqrt{g^{-1}f} = \{W, Y/a, -Y/a, -Y/a\}$ which would give rise to non-analyticity of the type which has been considered in our Subsection 5.1. The standard perturbation theory would not be well defined (compare to \cite{NBMG}).

In higher dimensions the concrete calculations become more and more involved. We don't aim at tackling this issues in the present paper. However, it seems reasonable to conjecture that the quadratic Lagrangian around Minkowski would always look like $$C_1\cdot\sum\limits_{i<j;\ i,j\in {\mathcal I}_{-}}\lambda_i\lambda_j+C_2\cdot\sum\limits_{i<j;\ i,j\in {\mathcal I}_{+}}\lambda_i\lambda_j$$ where $C_1$ and $C_2$ are constants, and ${\mathcal I}_{-}$ and ${\mathcal I}_{+}$ are the sets of indices corresponding to negative and positive eigenvalues of $\sqrt{\mathbb I}$ respectively.

\section{Discussion and Conclusions}
\label{Concl}

It is amazingly difficult to promote the simple action of Fierz and Pauli to a full-fledged non-linear theory of massive graviton. In the metric formulation one needs to compute the square root function of the matrix $g^{-1}f$. This procedure is definitely not among the simplest recipes of modern theoretical physics. And even in perturbation theory it is not so easy to deal with perturbative expansion of the square root, unless around a matrix proportional to identity. In some cases, it is not a mere technicality but rather a severe obstacle for correctly introducing small perturbations.

There is some freedom in taking the square root of a matrix. It can be of two types. The first type is absolutely generic. It gives discrete freedom which  basically refers to the usual two branches of the square root function in complex analysis. The peculiarity is that those branches can be independently chosen for different Jordan blocks. The second type is continuous and happens only in special cases. If we have taken different branches for Jordan blocks with equal eigenvalues, then a manifold of solutions arises given by those similarity transformations which change the square root without affecting the initial matrix.

In presence of continuous freedom, the naive perturbation theory in terms of matrices breaks down. Generic perturbation collapses the manifold to a finite number of points which depend on the chosen perturbation. Perturbation theory around such square roots is not well defined. A positive observation on top of that is that the similarity transformations which generate the manifold do not change the eigenvalues, and the eigenvalues are the only quantities which affect the Lagrangian. Of course, working directly with eigenvalues is very difficult. But fortunately everything can be done in terms of their elementary symmetric polynomials, and the mission is rendered accomplishable.

In this paper we have described the mathematical issues of working with square roots in massive gravity and developed the new perturbation theory in terms of eigenvalues and their elementary symmetric polynomials. The latter appeared to be a suitable tool for perturbation theory, even when around unusual roots. Our method allows to meaningfully define perturbations in cases of continuous freedom, even though with non-analyticity. Note that the origins of non-analyticity are very clear, however its implications remain to be understood, as well as its possible relations to the vielbein formulations of massive gravity.

\vspace{4ex}
{\Large \textbf{Acknowledgements}}

\vspace{2ex}
FS thanks Nicolay Gordeev and Yelena Yashina for useful discussions about some mathematical questions related to this paper. AG enjoyed many inspiring discussions about the square roots and other topics of massive gravity
with Fawad Hassan and Mikica Kocic. AG is grateful to the Dynasty Foundation and to the Saint Petersburg State University travel grant 11.42.687.2017 for support; and also
support from the Russian Science Foundation (project 16-11-10218) is gratefully acknowledged.

\begin{appendices}

\section{Some known facts from the theory of matrices}

In this Appendix we briefly review some well known facts from the theory of matrices over the field of complex numbers.

\subsection{Spectral theorem}

In finite-dimensional spaces one can give the full spectral classification of linear operators, without any requirements of self-adjointness or unitarity. In the language of matrices, an $n\times n$ matrix $\mathbb X$ can be brought by a similarity transformation to the Jordan normal form which is a block-diagonal form ${\mathcal S}{\mathbb X}{\mathcal S}^{-1}=J_{k_1}(\lambda_1)\bigoplus J_{k_2}(\lambda_2)\bigoplus\ldots$ with $k_i$ standing for the dimensionality of the $i$-th block, and the latter is given by
\[
J(\lambda)\equiv \begin{pmatrix}
\lambda & 1 & 0 & \cdots & 0 & 0 \\
0 & \lambda & 1 & \cdots & 0 & 0 \\        
\ldots & \ldots & \ldots & \ldots & \ldots & \ldots  \\
0 & 0 & 0 & \ldots &  \lambda & 1 \\
0 & 0 & 0 & \cdots & 0 & \lambda
\end{pmatrix}
\]
with the complex number $\lambda$ being the eigenvalue (of geometric multiplicity $1$ and algebraic multiplicity $k$ in this particular block). In components, we have non-zero elements $J_{i,i}=\lambda$ and $J_{i,i+1}=1$. If all Jordan blocks are $1$-dimensional then the matrix is diagonalisable. Note also that a similarity transformation can be viewed as simply changing the basis in the linear space in which the operator acts. The matrix $\mathcal S$ maps one basis onto another.

Various proofs of this theorem can be found in numerous textbooks on matrices. Here we only sketch a simple one. Given an $n\times n$ matrix $\mathbb X$ over complex numbers, one can always find an eigenvector\footnote{Recall that the solvability condition for the eigenvector equation $\mathbb Xe=\lambda e$ is $P(\lambda)=0$ where $P$ is the characteristic polynomial of $\mathbb X$. And the latter always has a solution in complex numbers.}, and choosing it as the $n$-th element of the basis we make ${\mathbb X}_{n,k}=0$ for $k<n$. After that we do the same for the upper left $(n-1)\times (n-1)$ corner of the new matrix, and so on. By induction, we bring $\mathbb X$ to an upper triangular form. Obviously, the numbers on the diagonal are eigenvalues, i.e. the roots of the characteristic polynomial of $\mathbb X$. By renumbering the elements of the basis, we can group them into sets of equal eigenvalues (if there are such). Let us choose one of the eigenvalues, $\lambda_1$. Then, we have $\mathbb X - \lambda_1 {\mathbb I}=\left(\begin{matrix}{\tilde X} & M\\ {\mathbb O} & X_0 \end{matrix}\right)$ where $\tilde X$ is non-degenerate and $X_0$ is nilpotent. If we make a similarity transformation by matrix $\left(\begin{matrix}{\mathbb I} & C\\ {\mathbb O} & \mathbb I \end{matrix}\right)$ then the matrix $\mathbb X - \lambda_1 {\mathbb I}$, and also the matrix $\mathbb X$, becomes block-diagonal if ${\tilde X}C-CX_0=M$ which can be solved as $C=\sum\limits_{i\geq 0}{\tilde X}^{-i-1}MX_0^i$ which is a well-defined finite sum since $\tilde X$ is non-degenerate and $X_0$ is nilpotent. By induction, we separate all blocks with different eigenvalues. Now we only have to understand the structure of a matrix $X_{\lambda_1}$ with all eigenvalues equal to a given number $\lambda_1$. Obviously, it suffices to study the nilpotent matrix $X_0=X_{\lambda_1}-\lambda_1\mathbb I$ of a given size $k\times k$. Actually, what we need now is a very generic statement which does not rely on the nice properties of the complex numbers: for any nilpotent linear operator in any (finite-dimensional) linear space one can decompose the vector space into direct sum of cyclic subspaces. Let's take the minimal number $k_1\leq k$ such that $X_0^{k_1}=0$. Then there is a vector $e_0$ for which $X_0^{k_1-1}e_0\neq 0$. This vector together with all vectors $X_0^i e_0$  generates a vector subspace $\mathcal V_0$ in which the matrix $X_0$ acts as a Jordan block with $\lambda=0$. If $k_1=k$, we are done. Otherwise, let's take a suitable\footnote{If the word "suitable" sounds as too sloppy a definition, one might think of the vector $e_1$ as one of those which require the maximal power of $\mathbb X$ to be sent to $\mathcal V_0$ (including zero). Or even more formally, one can consider the factor space of the full $\lambda=0$ block over $\mathcal V_0$, and perform the construction there. Since $\mathcal V_0$ is an invariant subspace, the action of $\mathbb X$ is correctly defined on the factor space, and any representative of the highest vector can serve as $e_1$.} vector $e_1$ from a linear complement to $\mathcal V_0$. We want to build a new cyclic subspace by vectors $X_0^i e_1$. This strategy might fail if for some $i$ we get $0\neq X_0^i e_1\in V_0$. Then it means that the vector $X_0^i e_1$ is a linear combination of vectors $X_0^j e_0$ with $j\geq i$ (for otherwise we have $X_0^{k_1}e_1\neq0$). Therefore, there is some $\tilde e \in \mathcal V_0$ such that $X_0^i \tilde e= X_0^i e_1$, and one can substitute $e_1$ by $e_1^{\prime}=e_1-\tilde e$ to overcome this difficulty. This procedure can be repeated if needed. By induction, we prove that $X_0$ can be brought to the form of direct sum of nilpotent Jordan blocks of sizes $k_i$ such that $\sum k_i=k$. The spectral theorem is proven.

\subsection{Analytic functions of matrices}

Similarity transformations preserve polynomial relations. If $P$ is a polynomial, then obviously $P({\mathcal S}{\mathbb X}{\mathcal S}^{-1})={\mathcal S}P(\mathbb X){\mathcal S}^{-1}$. Of course, it can be extended to any functions of matrices defined by norm-convergent series. The Jordan normal form gives also a very convenient tool for representing functions of matrices by
$$f\left({\mathcal S}^{-1}\left(\bigoplus_i J_{k_i}(\lambda_i)\right){\mathcal S}\right)={\mathcal S}^{-1}\left(\bigoplus_i f\left(J_{k_i}(\lambda_i)\right)\right)\mathcal S.$$

One can easily compute the powers of the Jordan block and see that for an analytic function $f$ which is given by its Taylor series it gives
\[
f(J_k(\lambda))\equiv \begin{pmatrix}
f(\lambda) & f^{\prime}(\lambda) & \frac12 f^{\prime\prime}(\lambda) & \cdots & \frac{1}{(k-2)!} f^{(k-2)}(\lambda) & \frac{1}{(k-1)!} f^{(k-1)}(\lambda) \\
0 & f(\lambda) & f^{\prime}(\lambda) & \cdots & \frac{1}{(k-3)!} f^{(k-3)}(\lambda) & \frac{1}{(k-2)!} f^{(k-2)}(\lambda) \\        
\ldots & \ldots & \ldots & \ldots & \ldots & \ldots  \\
0 & 0 & 0 & \ldots &  f(\lambda) & f^{\prime}(\lambda) \\
0 & 0 & 0 & \cdots & 0 & f(\lambda)
\end{pmatrix}.
\]
It allows to extend the definition to the class of $\mathcal C^{k-1} (\mathfrak U)$ functions where $k$ is the maximal size of the Jordan blocks, and $\mathfrak U \in \mathbb R$ is a domain which contains the full spectrum of the matrix.

In particular, since $f(x)=\sqrt{x}$ is smooth for $x\neq 0$, we have a definition of the square root of any non-degenerate matrix. Since the square root is a multivalued function, we see that the definition of $\sqrt{\mathbb X}$ admits the discrete freedom of choosing the sign of $\sqrt{\lambda}$ in each Jordan block. One has a matrix from the similarity class of $J_k(\pm\sqrt{\lambda})$ for $\sqrt{J_k(\lambda)}$ where $\lambda\neq 0$. We will see below that it is not the end of the story if we look at square roots as solutions of the matrix equation $\mathbb A\cdot \mathbb A=\mathbb X$.

Before doing that, let us comment on degenerate matrices. One can easily see that $$\left(\begin{matrix}0 & 1 & 0\\ 0 & 0 & 1\\ 0 & 0 & 0\end{matrix}\right)\cdot \left(\begin{matrix}0 & 1 & 0\\ 0 & 0 & 1\\ 0 & 0 & 0\end{matrix}\right) = \left(\begin{matrix}0 & 0 & 1\\ 0 & 0 & 0\\ 0 & 0 & 0\end{matrix}\right).$$
Therefore, $\left(J_3(0)\right)^2$ is equivalent to $J_2(0)\bigoplus J_1(0)$ under renumbering of the basis' elements. The Jordan block splits into two. Obviously, the splitting is related to the fact that $f^{\prime}(\lambda)=0$. Note that it never occurs if $f^{\prime}(\lambda)\neq 0$. 

Given also that $\left(J_2(0)\right)^2=\mathbb O$, we see that a $3\times 3$ matrix $\mathbb A$ such that ${\mathbb A}^2=J_3(0)$ does {\it not} exist. For some degenerate matrices, square roots do not exist at all, even over the field of complex numbers.

\subsection{Commuting matrices}

In order to determine all solutions of quadratic equation $\mathbb A\cdot \mathbb A=\mathbb X$, we will need to know the full set of solutions for commutation equation of the form $\mathbb A\mathbb X=\mathbb X\mathbb A$. 

Let us work in the basis in which the given matrix $\mathbb X$ acquires the canonical Jordan form $\mathbb X = \bigoplus\limits_i J_{k_i}(\lambda_i)$. This is a block diagonal form, and so let us consider the matrix $\mathbb A$ in a block form composed of $k_i\times k_j$ matrices $A_{i,j}$. Then the equation $\mathbb A\mathbb X=\mathbb X\mathbb A$ takes the form $A_{i,j}J_{k_j}(\lambda_j)=J_{k_i}(\lambda_i)A_{i,j}$ (no summation). 

If $\lambda_i\neq \lambda_j$ we get $(\lambda_i-\lambda_j)A_{i,j}=J_{k_i}(0)A_{i,j}-A_{i,j}J_{k_j}(0)$. The matrices $J(0)$ in the right hand side are nilpotent. We can multiply this equation by $\lambda_i - \lambda_j$ as many times as we want, and have higher powers of nilpotent matrices in the right hand side. Finally, we get $A_{i,j}=0$.

If $\lambda_i=\lambda_j$, then we have a simple equation $A_{i,j}J_{k_j}(\lambda)=J_{k_i}(\lambda)A_{i,j}$. The Jordan blocks have very simple structure, and one can easily check that if $k_i=k_j$ (for example, $i=j$ - diagonal blocks) then $A_{i,j}$ is an upper triangular matrix with components $\left(A_{i,j}\right)_{a,a+b}=C_b$ given by arbitrary numbers $C_b$ for $b=0, 1, \ldots k-1$. If $k_i\neq k_j$ then the solution is either $A_{i,j}=\left(\begin{matrix} \mathbb O & {\tilde A}_{i,j}\end{matrix}\right)$ or $A_{i,j}=\left(\begin{matrix} {\tilde A}_{i,j} \\
\mathbb O \end{matrix}\right)$ where ${\tilde A}_{i,j}$ is an upper triangular matrix of the same form as presented above and the size of ${\rm min}(k_i,k_j)\times {\rm min}(k_i,k_j)$.

If the matrix $\mathbb X$ was diagonalisable, then all the blocks are $1\times 1$-dimensional, and we simply have that the matrix element $A_{ij}$ is arbitrary if $\lambda_i=\lambda_j$, and $A_{ij}=0$ if $\lambda_i\neq \lambda_j$. It is very easy to understand: in those subspaces where the matrix $\mathbb X$ is proportional to identity it commutes with anything. And in general we need the two matrices to be {\it simultaneously} diagonalisable which is well known in quantum mechanics as the condition for two quantities to be simultaneously measurable.

\subsection{Square roots of matrices}

Let us now assume we are given a non-degenerate matrix 
$${\mathbb X}={\mathcal S}^{-1}\left(\bigoplus_i J_{k_i}(\lambda_i)\right){\mathcal S}$$
and need to solve the quadratic equation ${\mathbb A}^2=\mathbb X$ for $\mathbb A$. The matrix $\mathbb A$ can also be decomposed into the direct some of non-degenerate Jordan blocks which do not split under taking the square of the matrix. And therefore there is a one-to-one correspondence between the Jordan blocks of $\mathbb X$ and $\mathbb A$. And, obviously, instead of $J_k(\pm\sqrt{\lambda})$ we can use $\sqrt{J_k(\lambda)}$ in the spectral decomposition for $\mathbb A$:
$${\mathbb A}={\mathcal T}^{-1}\left(\bigoplus_i \sqrt{J_{k_i}(\lambda_i)}\right){\mathcal T}$$
where $\mathcal T$ gives a similarity transformation to be found, with the simplest solution being $\mathcal T = \mathcal S$.

Given that ${\mathbb A}^2=\mathbb X$ we have: ${\mathcal T}^{-1}\left(\bigoplus\limits_i J_{k_i}(\lambda_i)\right){\mathcal T}={\mathcal S}^{-1}\left(\bigoplus\limits_i J_{k_i}(\lambda_i)\right){\mathcal S}$. It easily translates into the commutation property:
${\mathcal S}{\mathcal T}^{-1}\left(\bigoplus\limits_i J_{k_i}(\lambda_i)\right)=\left(\bigoplus\limits_i J_{k_i}(\lambda_i)\right){\mathcal S}{\mathcal T}^{-1}$. One solution is always $\mathcal T=\mathcal S$, while the general solution is that $\mathcal T{\mathcal S}^{-1}$ commutes with $\bigoplus\limits_i J_{k_i}(\lambda_i)$. In other words, $\mathcal T$ is a product of $\mathcal S$ and an arbitrary matrix which commutes with $\bigoplus\limits_i J_{k_i}(\lambda_i)$. Of course, it gives a new square root only if $\mathcal T{\mathcal S}^{-1}$ does {\it not} commute with $\bigoplus\limits_i \sqrt{J_{k_i}(\lambda_i)}$.

Now, we can see where appears a continuous freedom in square roots. If there are two Jordan blocks in $\mathbb X$ with equal eigenvalues, $\lambda_i=\lambda_j$, then a similarity transformation matrix $\mathcal S$ which leaves $\mathbb X$ invariant can have non-trivial components in the $S_{i,j}$ block which were described above in A.3. At the same time, if we chose to have $\sqrt{\lambda_i}\neq\sqrt{\lambda_j}$, then such transformation {\it does} change the square root $\sqrt{\mathbb X}$. This is how we get the continuous freedom  corresponding to the manifold of similarity transformations which change $\sqrt{\mathbb X}$ leaving $\mathbb X$ intact.

\subsection{Cayley-Hamilton theorem}

Let $P(\lambda)$ be the characteristic polynomial of $\mathbb X$. The theorem states that $P(\mathbb X)=\mathbb O$. It trivially follows from our considerations above. Indeed, if there is a Jordan block $J_{k_i}(\lambda_i)$ in the spectral decomposition of $\mathbb X$, then it means that $\lambda_i$ is a root of $P$ of order not less than $k_i$. Therefore, $P(J_{k_i}(\lambda_i))=\mathbb O$ for every block in the spectral representation of $\mathbb X$.

\end{appendices}

\end{document}